\documentclass[sigconf]{acmart}

\usepackage{CJK}
\usepackage{multicol}
\usepackage{tabularx}
\usepackage{booktabs}
\usepackage{multirow}
\usepackage{subfigure}
\usepackage[T1]{fontenc}

\AtBeginDocument{%
  \providecommand\BibTeX{{%
    \normalfont B\kern-0.5em{\scshape i\kern-0.25em b}\kern-0.8em\TeX}}}

\copyrightyear{2020}
\acmYear{2020}
\setcopyright{acmcopyright}
\acmConference[SIGIR '20] {43rd International ACM SIGIR Conference on Research and Development in Information Retrieval}{July 25--30, 2020}{Virtual Event, China}
\acmBooktitle{43rd International ACM SIGIR Conference on Research and Development in Information Retrieval (SIGIR '20), July 25--30, 2020, Virtual Event, China}
\acmPrice{15.00}
\acmDOI{10.1145/3397271.3401324}
\acmISBN{978-1-4503-8016-4/20/07}

\begin{document}

\title{ User-Inspired Posterior Network \\for Recommendation Reason Generation }


\def\first{$^1$}
\def\second{$^2$}
\def\third{$^3$}
\def\forth{$^4$}
\def\fifth{$^5$}
\def\comma{$^,$}
\def\star{$^*$}

\author{{Haolan Zhan}, Hainan Zhang, Hongshen Chen, Lei Shen, Yanyan Lan, Zhuoye Ding, Dawei Yin}
\authornote{Work done at Data Science Lab, JD.com. \\ Corresponding Authors: Hainan Zhang and Hongshen Chen.}
\affiliation{
\institution{{State Key Lab. of Computer Science, Institute of Software, Chinese Academy of Sciences}}}
\affiliation{
\institution{{Institute of Computing Technology, Chinese Academy of Sciences}}}
\affiliation{
\institution{{University of Chinese Academy of Sciences}}}
\affiliation{
\institution{{Data Science Lab, JD.com;  Baidu Inc.}}}

\email{zhanhaolan316@gmail.com, zhanghainan@jd.com, ac@chenhongshen.com}
\email{{shenlei17z, lanyanyan}@ict.ac.cn, dingzhuoye@jd.com, yindawei@acm.org}

\fancyhead{}
\begin{abstract}

Recommendation reason generation, aiming at showing the selling points of products for customers, plays a vital role in attracting customers' attention as well as improving user experience. A simple and effective way is to extract keywords directly from the knowledge-base of products, i.e., attributes or title, 
as the recommendation reason.
However, generating recommendation reason from product knowledge doesn't naturally respond to users' interests. 
Fortunately, on some E-commerce websites, there exists more and more user-generated content (user-content for short), i.e., product question-answering (QA) discussions, which reflect user-cared aspects.
Therefore, in this paper, we consider generating the recommendation reason by taking into account not only the product attributes but also the customer-generated product QA discussions.
In reality, adequate user-content is only possible for the most popular commodities, whereas large sums of long-tail products or new products cannot gather a sufficient number of user-content.
To tackle this problem, we propose a user-inspired  multi-source posterior transformer (MSPT), which induces the model reflecting the users' interests with a posterior multiple QA discussions module, and generating recommendation reasons containing the product attributes as well as the user-cared aspects.
Experimental results show that our model is superior to traditional generative models.  
Additionally, the analysis also shows that our model can focus more on the user-cared aspects than baselines.

\end{abstract}
\vspace{-5mm}





\maketitle

\section{Introduction}

\begin{figure}[!t]
\centering

\subfigure[Screenshot of JD App.]{
\begin{minipage}[t]{0.5\linewidth}
\centering
\includegraphics[width=0.8\linewidth]{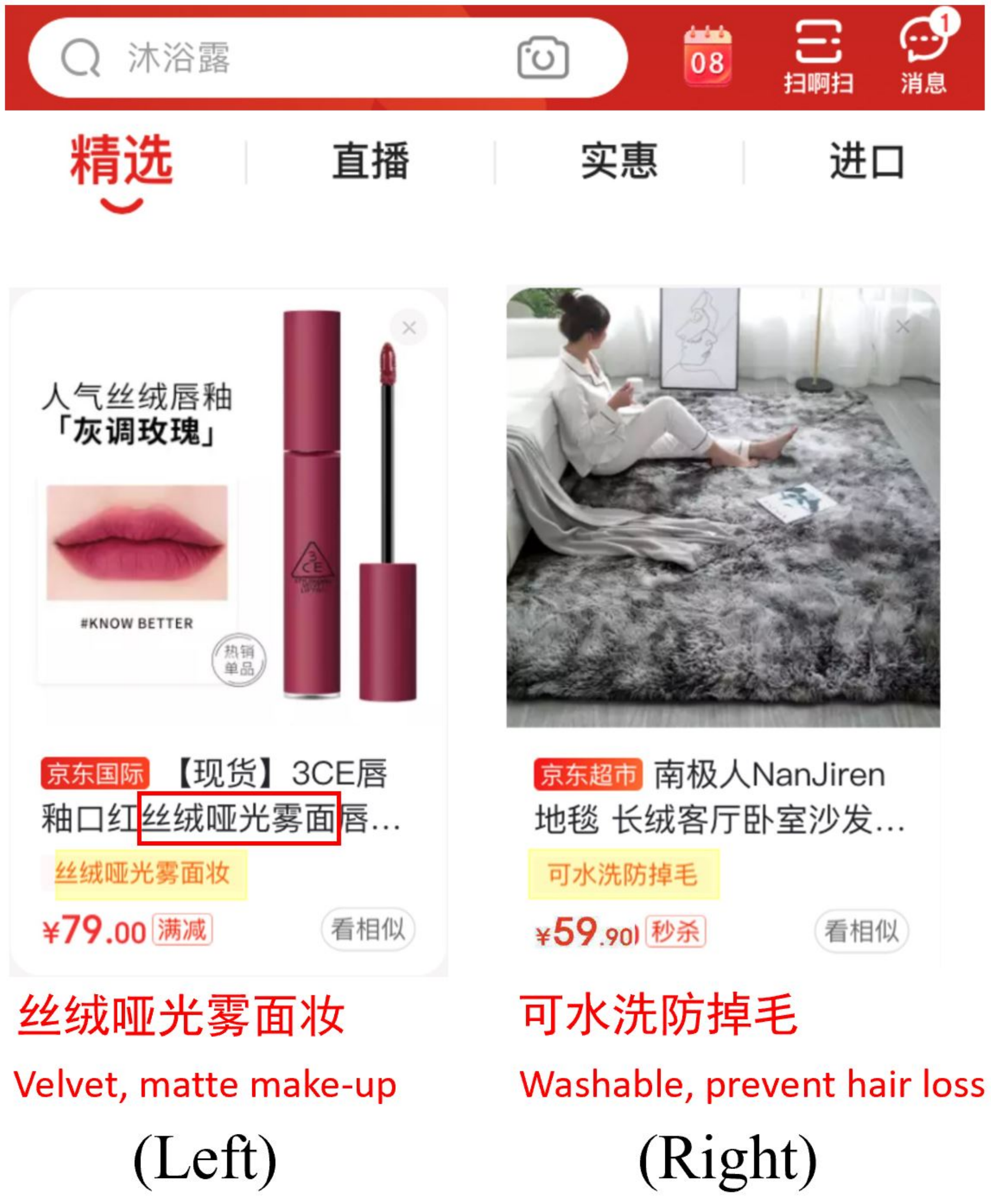}
\label{screenshot}
\end{minipage}
}%
\subfigure[QA discussions of the right product.]{
\begin{minipage}[t]{0.5\linewidth}
\centering
\includegraphics[width=0.8\linewidth]{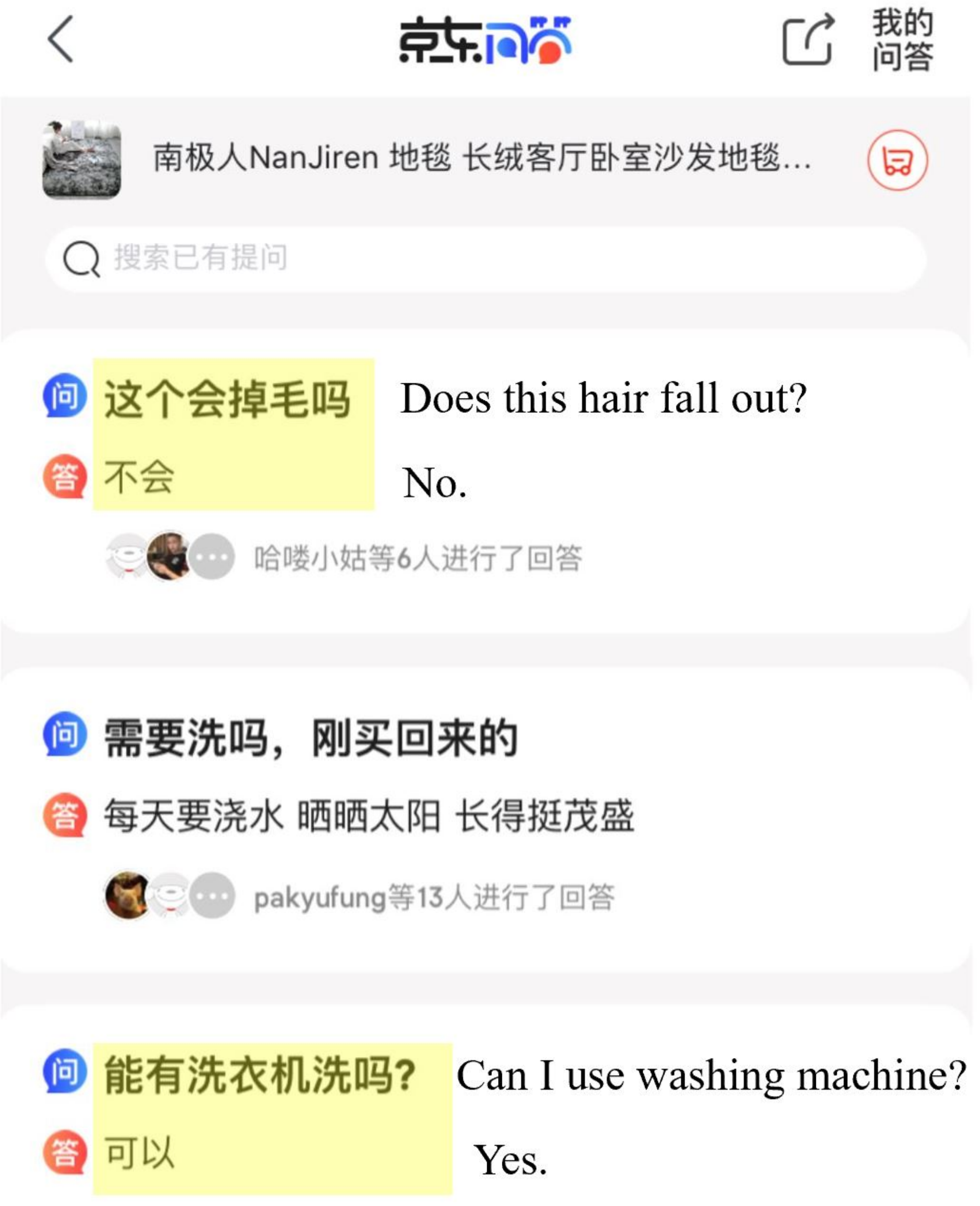}
\label{qa_case}
\end{minipage}
}%

\centering
\caption{\small{Two products with title and recommendation reason.}}
\vspace{-5mm}
\label{fig:intro_case}
\end{figure}


Nowadays, more and more customers are used to online-shopping based on the recommendation system rather than just search for a product directly.
However, existing recommendation systems, just ranking the products, can no longer meet the requirements of customers. To save costs, most customers would like to browse products with the selling points in the limited mobile feeds. Thus, it is necessary to obviously display the selling points as the recommendation reasons to help customers learn the recommended products straightforwardly.

Furthermore, products with informative recommendation reasons can also attract customers’ attention and receive more clicks. From screenshots in Figure \ref{screenshot}, we can see that the red line, showing the informative characteristics of the product, i.e., ``\begin{CJK*}{UTF8}{gkai}{可水洗防掉毛 (Washable, prevent hair loss)}\end{CJK*}'', can arouse customers' interests and encourage them to click to see the details. Based on our online analysis, we find that products with recommendation reasons can receive more clicks than those without. 
Such effects of informative display have also been discussed in previous studies\cite{sun2018multi,zhang2019multi}, showing that when generated properly, the recommendation reasons will increase the profits of the recommendation system.

Given the product information, such as title, attributes and relevant QAs, RRG task aims at generating selling points of the product. 
The explicit approach can be inspired by the Short Title Generation (STG) task \cite{sun2018multi} for E-commerce, which automatically generates short titles by extracting key information from item-content, i.e., product title and its attributes. 
The researchers propose to use the pointer-network to copy the brand name and the key information from the product title and attributes as the short title.
Similarly, RRG can be viewed as a modified form of the STG, with the emphasis on creating a natural reason from the product item-content by taking the key information as selling points.
As shown in Figure \ref{screenshot}, the example ``\begin{CJK*}{UTF8}{gkai}{丝绒哑光面妆 (Velvet, matte makeup)}\end{CJK*}'' is an item-content based recommendation reason. It is generated by extracting the selling points directly from title and attributes, which can be seen as an important and explicit method for the RRG task.

However, this item-content based approach is not concerned with the user's experience and the most-cared questions, which may hurt the effectiveness of the generated reason.
Obviously, the RRG should take more respect to the user-cared aspects rather than the item-content.
Therefore, how to incorporate user-cared aspects into the recommendation reason is critical for RRG.
To tackle this problem, we propose to use the user-content (QA discussions) as the most-cared aspects from the user's experience. Taking the right product in Figure~\ref{qa_case} as an example, some customers post their questions about the product and others answer these questions. The recommendation reason can be summarized as ``\begin{CJK*}{UTF8}{gkai}{可水洗不掉毛 (Washable, prevent hair loss)}\end{CJK*}''. Therefore, the most-cared aspect of the product can be mined from the user-content.

In reality, adequate user-content is only possible for the most popular commodities, whereas large sums of long-tail products or new products can not gather sufficient amount of product user-content. 
In this paper, we propose a novel model named Multi-Source Posterior Transformer (MSPT) to explicitly model these two aspects and take into account the user-content with an adaptive posterior module.
Specifically, for the item-content, we first use the Bi-LSTM encodes the title and attributes, and then use the self-attention \cite{vaswani2017attention} to obtain the title representation and the attribute representation. Besides, for the user-content, we utilize the product-guided attention module to encode the QA sentences into dense representation with a  hierarchical method. Thirdly, we train the posterior network with feature fusion between QA pairs and products through minimizing the KL divergence of prior and posterior distributions. Finally, we use the posterior representation as input for decoding phase.

For experiments, we construct a large-scale JDRRG dataset containing 122,405 product items form JD App.
We compare our model with several state-of-the-art generation baselines using both automatic and human evaluations.
The results demonstrate that our model performs significantly better than existed methods. 

\vspace{-2mm}

\section{Problem Statement} \label{sec:data}



\textbf{Task Definition:}
Given a product title, attribute words and its relative question answering pairs, the RRG task aims to utilize inherent product information to identify the privilege of products and users' preference, as well as to generate the coherent and appealing recommendation reasons. 
Formally, given a product title $T$ composed of a sequence of words $\{t_1,...,t_N\}$, a set of attribute words $W=\{w_1,...,w_M\}$ and a relative question-answering set defined as $\{\{Q_1,A_1\},...,\{Q_n,A_n\}\}$, the RRG task is to learn the generative model $G(\cdot)$. Each question in $Q$ is defined as $Q_n=\{q_n^1,\cdots,q_n^L\}$, where $q_n^i$ is the $i$-th word in the sentence of $Q_n$ and $L$ is the max length of sentence. Each answer in $A$ is defined as $A_n=\{a_n^1,\cdots,a_n^L\}$, where $a_{n}^j$ is the $j$-th word in the answer sentence. The corresponding recommendation reason is defined as $Y=\{y^1,\cdots,y^S\}$, where $y^i$ is the $i$-th word and $S$ is the max length. 


\textbf{Dataset Collection:}
As there is no appropriate corpus for our RRG task currently, we build a new Chinese dataset, JDRRG. We collect our dataset from JD App, a Chinese E-commerce website. E-commerce is a popular way of shopping, where customers can post questions about products in public and other customers can give reply answers. JDRRG dataset consists of 122,405 products and relative 3.79 million pairs of QA discussions from real customers and 122,405 annotated recommendation reasons from our invited annotators (all professional writers). Each product has 30.9 different QA pairs on average. The average length of each question sentence, answer sentence, the product title, attributes and recommendation reason are 18.05, 18.59, 23.2, 16.03 and 10.5 respectively.
\section{Proposed Model}


\begin{figure}[!t]
\begin{center}
   \includegraphics[width=1.0\linewidth]{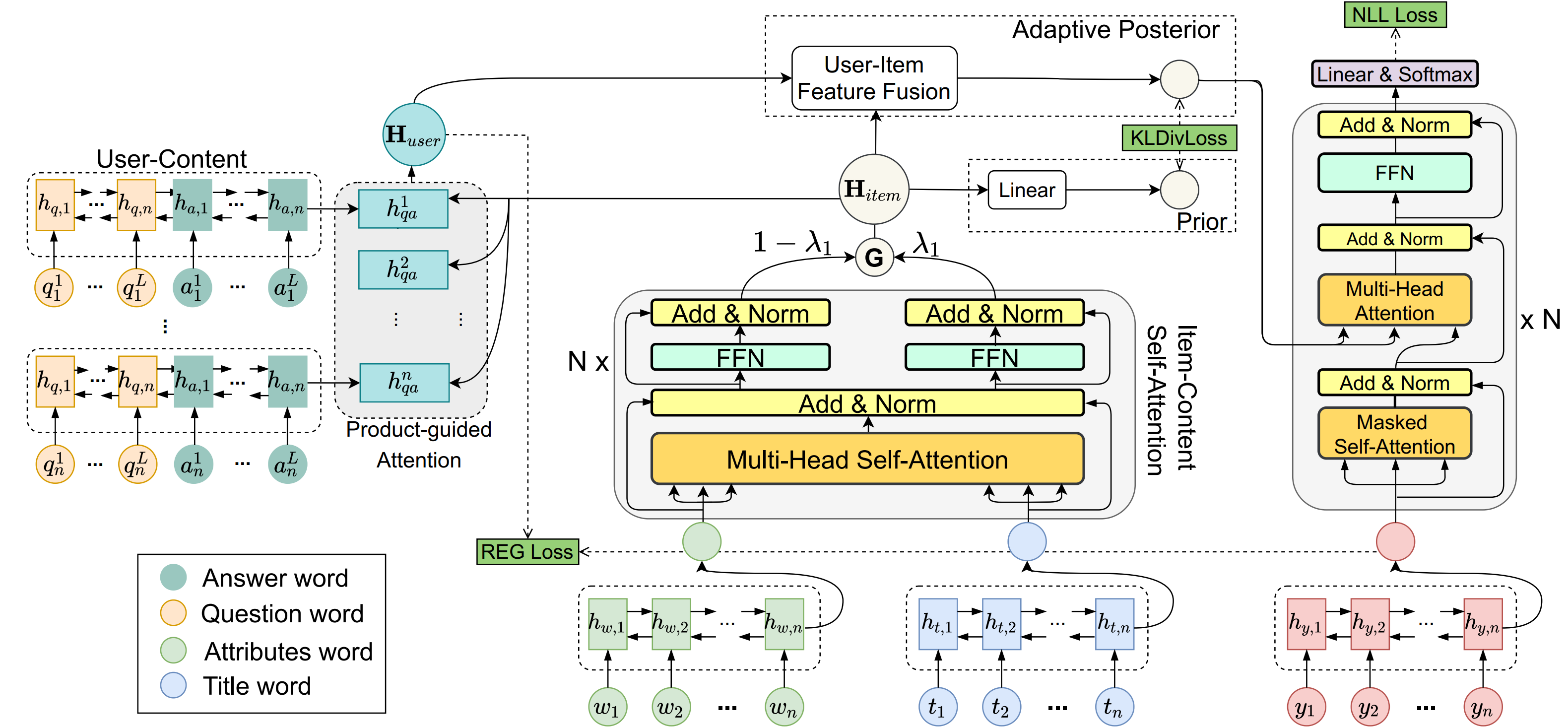}
\end{center}
    \caption{The framework of our proposed Multi-Source Posterior Transformer Network (MSPT).}
\label{fig:framework}
\vspace{-5mm}
\end{figure}

\subsection{Item-content Self-Attention Module}
As shown in Figure~\ref{fig:framework}, our proposed MSPT model consists of an item-content self-attention module, a user-content QA attention module, an adaptive posterior module followed with decoding phase.

\subsubsection{Word-level Encoder}\label{sec:lstm}

Given a product title $T$$=$$\{t_1,\dots,t_N\}$ as the input, a standard Bi-LSTM first maps the input sequence to a fixed-dimension vector $\mathbf{h}_T={[h_L,h_R]}$.
At the same time, we also use another Bi-LSTM to encode the attributes $W$ as $\mathbf{h}_W$.

\subsubsection{Title and Attributes Self-Attention}\label{sec:product}
 Given the word-level representation $\mathbf{h}_T$ and $\mathbf{h}_W$ of title and attributes, the initial  input of the self-attention \cite{vaswani2017attention} is the sum of the word-level representation and position encoding (PE), denoting as $ PE(\mathbf{h}_{T})$ and $PE(\mathbf{h}_{W})$, respectively.
$\mathbf{E}^{(0)}_{T}$ and $\mathbf{E}^{(0)}_{W}$ are the initial input representations at the first layer. Self-attention is a special attention mechanism in a normal Transformer architectures \cite{vaswani2017attention}. At the $l$-th layer, the output representation is defined as belowed:
\begin{align}
    \mathbf{E}_{T}^{(l)} &= {\rm FFN}({\rm MHA}(\mathbf{E}_{T}^{(l-1)}, \mathbf{E}_{T}^{(l-1)}, \mathbf{E}_{T}^{(l-1)})),\\ 
    \mathbf{E}_{W}^{(l)} &= {\rm FFN}({\rm MHA}(\mathbf{E}_{W}^{(l-1)}, \mathbf{E}_{W}^{(l-1)}, \mathbf{E}_{W}^{(l-1)})),
\end{align}
where $\mathbf{E}_{T}^{(l)}$, $\mathbf{E}_{W}^{(l)}$ denote the output representations after the $l$-th layer. The sub-layer ${\rm FFN}$ is a position-wise fully connected feed-forward network and ${\rm MHA}(\mathbf{Q},\mathbf{K},\mathbf{V})$~\cite{vaswani2017attention} is a multi-head attention function.
The final product representation is defined as:
\begin{align}
    \mathbf{H}_{item}=\lambda_{1}\mathbf{E}_{T}^{(N)}+(1-\lambda_{1})\mathbf{E}_{W}^{(N)},
\end{align}
where $\lambda_{1}$ is a parameter, and $\mathbf{E}^{(N)}_{T}$ and $\mathbf{E}^{(N)}_{W}$ is the final representation of title and attributes that output from the $N$-th layer.

\subsection{User-content QA Attention Module}
The user-content QA attention module
consists of the word-level encoder for each QA sentence and context-level encoder for multiple QA pairs with the product-guided attention. The product-guided attention (PGA) examines all the relationships to highlight the most relevant QA pairs to the product. This type of attention is guided by the current product representation which is enhanced with corresponding information from the QA discussions.
For each question sentence $Q_n$ and answer sentence $A_n$ in product a given product, we concatenate both sentences as input, and use the Bi-LSTM encoder to obtain its low-dimension representation $\mathbf{h}_{QA}^{n}$, the same in section~\ref{sec:lstm}.
Given the QA representation $\mathbf{h}_{QA}=\{\mathbf{h}_{QA}^{1},\mathbf{h}_{QA}^{2},\dots,\mathbf{h}_{QA}^{n}\}$,
the context-level representation for multiple QA pairs is defined as a weighted sum of these $\mathbf{h}_{QA}^k$:
\begin{equation}
    \mathbf{H}_{user}=\sum\nolimits_{k=1}^{n}{\delta_{k}^{qa}\mathbf{h}_{QA}^k}.
\end{equation}
where weight $\delta_{k}^{qa}$ of each representation $\mathbf{h}_{QA}^k$ is defined as:
\begin{equation}
\delta_{k}^{qa} = softmax((\mathbf{W}_{1}\mathbf{H}_{item} + \mathbf{b}_{1})^{T}(\mathbf{W}_{2}\mathbf{h}_{QA}^{k} + \mathbf{b}_{2})).
\end{equation}
where $\mathbf{W}_{1}$,$\mathbf{b}_{1}$,$\mathbf{W}_{2}$ and $\mathbf{b}_{2}$ are learned parameters. $\mathbf{H}_{item}$ is the output of the item-content self-attention module in section~\ref{sec:product}. 
we obtain the final representation of the multiple QA discussion $\mathbf{H}_{user}$.

\subsection{Adaptive Posterior Module}
Given the user-content representation $\mathbf{H}_{user}$ and item-content representation $\mathbf{H}_{item}$, the goal of adaptive posterior module is to enhance our model to learn the user-cared aspect from QA discussions. The adaptive posterior module consists of two sub-modules: a prior module and a posterior  module.
In the prior module, the product representation is updated by a linear transform module:
\begin{equation}
    \widetilde{\mathbf{H}}_{prior} = \sigma(\mathbf{W}_{3}\mathbf{H}_{item} + \mathbf{b}_{3}).
\end{equation}

In the posterior module, we define the posterior representation by adopting the user-item feature fusion mechanism. We concatenate the item-content representation $\mathbf{H}_{item}$ and the user-content representation $\mathbf{H}_{user}$ at first. And then, we adopt a linear transform to get the final representation, named hard fusion:
\begin{equation}
    \widetilde{\mathbf{H}}_{post} = \sigma(\mathbf{W}_{4}[\mathbf{H}_{item}, \mathbf{H}_{user}] + \mathbf{b}_{4}).
\end{equation}

Besides, we also provide a soft fusion choice while employing the gate mechanism to fuse $\mathbf{H}_{item}$ and  $\mathbf{H}_{user}$.
\begin{align}
     \widetilde{\mathbf{H}}_{post} &= \sigma(\mathbf{W}_{5}(\lambda_{2}\mathbf{H}_{item}+(1-\lambda_{2})\mathbf{H}_{user}) + \mathbf{b}_{5}). 
\end{align}

Clearly, the discrepancy between prior and posterior distributions introduces great challenges in training the model. 
To tackle this problem, we introduce the KL divergence loss, to measure the proximity between the prior representation and the posterior representation. The KL divergence is defined as follows:
\begin{equation}
   \mathcal{L}_{KL}(\theta)= D_{KL}(p({y_{t}}|\widetilde{\mathbf{H}}_{post}) || p({y_{t}}|\widetilde{\mathbf{H}}_{prior});\theta),
\end{equation}
where $\theta$ denotes the model parameters.

\subsection{Training Objectives}


We employ three different loss functions: KL divergence,  NLL and Regular loss. All loss functions are also elaborated in Figure~\ref{fig:framework}.

\textbf{NLL Loss.} The objective of NLL loss is to quantify the difference between the ground truth reason and the generated reason by our model. It minimizes Negative Log-Likelihood (NLL):
\begin{equation}
   \mathcal{L}_{NLL}(\theta) = -\sum_{t=1}^{S} {\rm log} p(y_t|y_{<t}, T, W, \{{Q}_i,{A}_i\}_{i=1}^{n};\theta),
\end{equation}
where $\theta$ denotes the model parameters and $y_{<t}$ denotes the previously generated words.

\textbf{Regular Loss.} 
It aims to enforce the relevancy between the QA discussion representation and the
true reason. Then, the Regular loss is defined to minimize:
\begin{equation}
   \mathcal{L}_{REG}(\theta) = -\sum_{t=1}^{S} {\rm log} p(y_t|\mathbf{H}_{user};\theta).
\end{equation}

our final loss is a combination of these three loss functions:
\begin{equation}
   \mathcal{L}(\theta) = \mathcal{L}_{KL}(\theta)  + \mathcal{L}_{NLL}(\theta) + \mathcal{L}_{REG}(\theta).
\end{equation}


\section{Experiments}

\begin{table}[!t]
\small
\centering
\begin{tabular}{lcccccccc}
\toprule[1pt]
  Model & RG-1 & RG-2 & RG-L & BU-1 & BU-2 & MTR\\
  \hline
  TextRank & 8.04 & 2.19 & 7.32 & 7.89 & 1.74 &  2.03 \\
  SEQ2SEQ & 10.60 & 2.86 & 8.43 & 8.38 & 1.91 &   2.35 \\
  PG-BiLSTM & 12.32 &  3.59 & 9.35 & 10.54 & 2.35 &  3.25 \\
  MS-Ptr & 13.45 & 4.41 & 10.65 & 11.24& 2.69 &  3.98 \\ \hline
  Transformer & 11.21 & 3.46 & 9.68 & 9.89 & 2.01 &  3.47 \\
  CopyTrans & 12.37 & 4.39 & 10.50 & 10.95 & 2.24 &  4.04 \\
  HierTrans & 11.94 & 3.80 & 9.96 & 10.58 & 2.36 & 3.94 \\
  Hi-MAP & 13.85  & 4.65 & 11.48 & 11.03 & 2.85 &  4.31\\

  \midrule[1pt]
 {MSPT-\textit{QA}} & 13.54 & 4.23 & 10.59 & 10.94 & 2.47 &  3.93 \\
  {MSPT-\textit{PGA}} & 15.03 & 4.89 & 12.24 & 12.06 & 2.84 & 4.53\\
  \textbf{MSPT} (\textit{hardFS}) & 15.97 & 5.62 & 12.76 & 13.13 & 3.26 &  4.97\\
  \textbf{MSPT} (\textit{softFS}) & \textbf{16.52} & \textbf{5.83} & \textbf{13.38} & \textbf{15.69} & \textbf{3.52}  & \textbf{5.16}\\
 \bottomrule[1pt]
\end{tabular}
\caption{Metric-based evaluation on JDRRG dataset (\%).}
\label{tab_auto-eval}
\vspace{-5mm}
\end{table}


\subsection{Experimental Settings}

\textbf{Baseline Approaches:} 
\textbf{TextRank} is an extraction method using a graph-based ranking algorithm \cite{mihalcea2004textrank}. \textbf{SEQ2SEQ} is a basic end-to-end model based on standard LSTM units and the attention mechanism \cite{tan2015lstm}. 
\textbf{PG-BiLSTM} is  the bi-directional LSTM with pointer generator mechanism \cite{see2017get}. \textbf{MS-Ptr} represents the multi-source pointer network for short product title generation \cite{sun2018multi}.  \textbf{CopyTrans} is the Transformer model with the copy mechanism \cite{gehrmann2018bottom}.
\textbf{HierTrans} represents the hierarchical transformer for multi-document summarization (MDS) tasks \cite{liu-lapata-2019-hierarchical}. \textbf{Hi-MAP} is the hierarchical network with MMR-self-attention mechanism for MDS task \cite{fabbri-etal-2019-multi}.

\noindent\textbf{Model Variants:} 
To evaluate the product-guided attention module, User-Item feature fusion module and posterior module, we also employ some degraded MSPT models to investigate the effect of our proposed mechanisms: \textbf{MSPT}-QA removes the QA discussions and relevant product-guided attention and posterior module during training process.
\textbf{MSPT}-PGA is MSPT model without the product-guided attention mechanism in user-content QA attention module.
\textbf{MSPT} (\textit{hardFS}) is the MSPT model with hard feature fusion.
\textbf{MSPT} (\textit{softFS}) is the MSPT model with soft feature fusion.



\begin{table}[!t]
\small
\centering
\begin{tabular}{ccccc}
  \toprule[1pt] 
  \multirow{2}{0.9cm}{Model} & \multicolumn{3}{c}{MSPT vs.} & \multirow{2}{0.7cm}{kappa}\\ \cline{2-4}
  ~ & Win & Loss & Tie & ~\\
  \hline
  PG-BiLSTM & 57\% & 14\% & 29\% & 0.475 \\
  MS-Ptr & 43\% & 21\% & 36\% & 0.491 \\
  CopyTrans & 46\% & 28\% & 26\% & 0.427\\
  HierTrans & 51\% & 19\% & 30\% & 0.528\\
  Hi-MAP & 40\% & 32\% & 28\% & 0.443 \\
  \bottomrule[1pt]
\end{tabular}
\caption{Human evaluation of MSPT on JDRRG dataset.}
\label{tab:human}
\vspace*{-5mm}
\end{table}

\begin{table}[!t]
\footnotesize
    \centering
    \begin{tabular}{p{30pt}|p{190pt}}
    \toprule[1pt]
      \small{Product} &  \begin{CJK*}{UTF8}{gkai}\small{雅邦荔枝冻金箔果冻口红不易掉色唇膏女士口红}  \end{CJK*} \\
      \small{Title} &  \small{Alobon lychee gold foil jelly, not-easy-to-fade lady lipstick} \\ \hline \small{Attributes} &  \begin{CJK*}{UTF8}{gkai}\small{金箔,布丁,口红,保湿 (gold foil, jelly, lipstick, moisturizing)}  \end{CJK*} \\ \hline
      \small{QA Pair 1} & \begin{CJK*}{UTF8}{gkai}\small{Q: 保湿怎样？容易掉色吗？$\rightarrow$ A1: 保湿效果不错，我朋友也买了。/// A2: 不掉色，不沾杯。 /// A3:挺滋润的}  \end{CJK*} \\
      ~ & \small{Q: What about preserving moisture? Is the lipstick easy to fade? $\rightarrow$ A1: Pretty good, good moisturizing effect, my friends also bought it. /// A2: Don't fade, not sticky. /// A3: Great, it's moist.} \\ \hline
        \small{QA Pair 2} & \begin{CJK*}{UTF8}{gkai}\small{Q: 这款口红有什么味道的？ $\rightarrow$ A1: 淡淡的水果味，挺好闻。/// A2: 樱桃味，很不错。}  \end{CJK*} \\
        & \small{Q: What kinds of smells does this lipstick have? $\rightarrow$ A1: Light fruit flavor, it smells good. /// A2: Cherry flavor, Great.}  \\ 
        \midrule[1pt]
        \small{MS-Ptr} & \begin{CJK*}{UTF8}{gkai}\small{口感很清脆,精心礼盒装}  \end{CJK*}  \small{Very crisp taste, well-crafted gift pack} \\ \hline
        \small{HierTran} & \begin{CJK*}{UTF8}{gkai}\small{持久不掉色,清爽不油腻}  \end{CJK*}  \small{Lasting, refreshing, non-greasy}  \\ \hline
        \small{Hi-MAP} & \begin{CJK*}{UTF8}{gkai}\small{保湿滋润不掉色，包装不错易保湿}\end{CJK*} \small{Moisturizing, no fading, good packaging, easy to moisturize}\\ \hline
        \small\textbf{MSPT} & \begin{CJK*}{UTF8}{gkai}\small{保湿润唇不掉色，水果口味质量佳}  \end{CJK*} \small{Moisturizing for lips, does not fade, fruit scent, and good quality}  \\
    \bottomrule[1pt]
    \end{tabular}
    \caption{Case study of MSPT and baselines on JDRRG.}
    \label{tab_finalcase}
    \vspace{-3mm}
\end{table}

\subsection {Results and Analysis}

\textbf{Metric-based Evaluation.}
The metric-based evaluation results are shown in Table \ref{tab_auto-eval}. 
Our proposed MSPT model outperforms the best. Taking the ROUGE metrics as an example, the ROUGE-1 value of the MSPT is 16.52, which is significantly better than MS-Ptr, CopyTrans and Hi-MAP, i.e., 13.45, 12.37 and 13.85. The METEOR metrics of our model is also higher than other baseline models, indicating that our model can generate more informative and fluency recommendation reason. We also conducted a significant test, showing that the improvement is significant, i.e., $\text {p-value} < 0.01$. 

\noindent\textbf{Human Evaluation.}
The results of the human evaluation are shown in Table \ref{tab:human}. The percentage of win, loss, and tie, as compared with the baseline models, is shown to evaluate the relevance and the informativeness of the generated reasons by MSPT. As showed in \ref{tab:human}, the percentage of win is greater than the loss, indicating that our MSPT model is  better than all baseline methods. MSPT achieves a preference gain (i.e., the win ratio minus the loss ratio) of 43\%, 21\%, 18\%, 22\% and 18\% respectively, compared to PG-BiLSTM, MS-Ptr, CopyTrans, HierTrans and Hi-MAP, indicting that MSPT model can generate not only relevant but also informative reasons than baseline models. The Fleiss' kappa \cite{fleiss1971measuring} value demonstrates the consistency of different annotators. 

\noindent\textbf{Ablation Study.}
As shown in Table~\ref{tab_auto-eval}, we also conduct an ablation study with model variants. Above all, to validate the enhancement of QA discussions, we remove this module and get the MSPT-\textit{QA} model. The metric evaluation shows that the QA discussions perform very importantly. Besides, we also remove the product-guided attention module to get MSPT-\textit{PGA} model. It also lost in the competition with MSPT(\textit{soft}). Therefore, we can see that the product title and attributes have a good effect on selecting the relevant information in QA discussions. Moreover, the different feature fusion mechanism also makes a difference. The MSPT model with soft fusion performs better than the model with hard fusion. 



\noindent\textbf{Case Study.}
To facilitate a better understanding of our model, we present some examples in Table \ref{tab_finalcase}. We select three baseline models: MS-Ptr, HierTrans and Hi-MAP, since these models outperform other baselines in metric-based evaluation. As shown in \ref{tab_finalcase}, our proposed MSPT model generates more aspects of product with considering QA discussions. 
For example, the recommendation reason generated by HierTrans and MS-Ptr, such as ``\begin{CJK*}{UTF8}{gkai}{精心礼盒装 (well-crafted gift pack)}\end{CJK*}'' and ``\begin{CJK*}{UTF8}{gkai}{清爽不油腻 (Refreshing and non-greasy)}\end{CJK*}'', contains some of the incorrect information.  The  Hi-MAP can generate relevant and appropriate recommendation reason such as ``\begin{CJK*}{UTF8}{gkai}{保湿滋润不掉色 (Moisturizing, no fading)}\end{CJK*}''. However, this reason sentence contains repetitive information, such as ``\begin{CJK*}{UTF8}{gkai}{保湿 (Moisturize)}\end{CJK*}''. Our proposed MSPT model contains more characteristics, i.e., ``\begin{CJK*}{UTF8}{gkai}{水果口味 (fruit scent)}\end{CJK*}'' and ``\begin{CJK*}{UTF8}{gkai}{不掉色 (does not fade)}\end{CJK*}'', since it utilizes the QA discussions and enhances the user-cared aspects.


\section{Conclusion and Future work}

In this paper, we proposed a challenging recommendation reason generation task in E-commerce. To tackle the problem, we proposed a novel multiple source posterior transformer to incorporate user-cared aspects into the recommendation reason. Besides, we constructed a new
benchmark dataset JDRRG to study and evaluate this task. The experimental results demonstrated that our model could outperform baselines. In future work, we plan to further investigate the proposed model with emotion classification.


\bibliographystyle{ACM-Reference-Format}
\bibliography{sigir-fuben}

\end{document}